\documentclass[sigconf,screen]{acmart}
\setcopyright{cc}
\setcctype{by}
\acmDOI{10.1145/3832783.3837415}
\acmYear{2026}
\copyrightyear{2026}
\acmISBN{979-8-4007-2882-2/2026/10}
\acmConference[ASE '26]{Proceedings of the 41st IEEE/ACM International Conference on Automated Software Engineering}{October 12--16, 2026}{Munich, Germany}
\acmBooktitle{Proceedings of the 41st IEEE/ACM International Conference on Automated Software Engineering (ASE '26), October 12--16, 2026, Munich, Germany}
\acmSubmissionID{ase26main-p92-p}
\received{2026-03-26}
\received[accepted]{2026-06-18}

\AtBeginDocument{%
  
}

\usepackage[inline]{enumitem}
\usepackage{booktabs}

\usepackage{amsmath,amssymb}
\usepackage{algorithm}
\usepackage{array}
\usepackage{algpseudocode}
\usepackage{graphicx}
\usepackage{textcomp}
\usepackage[dvipsnames]{xcolor}
\usepackage{minted}
\usepackage{tabularx}
\usepackage{listings}
\usepackage{color}
\usepackage{multirow}
\usepackage[T1]{fontenc}
\usepackage{mathtools}
\usepackage{hyperref}
\usepackage{lipsum}
\usepackage{subcaption}
\usepackage{tcolorbox}
\usepackage{anyfontsize}
\usepackage{mdframed}
\usepackage{tikz}
\usepackage{makecell}
\usepackage[normalem]{ulem}
\usetikzlibrary{shapes.geometric, arrows.meta, positioning, fit, calc, backgrounds}

\usepackage{float}
\definecolor{dkgreen}{rgb}{0,0.6,0}
\definecolor{gray}{rgb}{0.5,0.5,0.5}
\definecolor{mauve}{rgb}{0.58,0,0.82}
\definecolor{mygray}{rgb}{0.9,0.9,0.9}
\DeclareUnicodeCharacter{2212}{-}

\makeatletter
\AtEndPreamble{%
  \global\@ACM@balancefalse
  \RequirePackage{pbalance}
}
\makeatother
\begin{document}

\title{A Unified Model for Cross-Domain Clone Detection via Model Merging}

\author{Palash R. Roy}
\correspondingauthor
\orcid{0000-0001-9470-4233}
\affiliation{%
  \institution{University of Saskatchewan}
  \department{Computer Science}
  \city{Saskatoon}
  \country{Canada}
}
\email{palash.roy@usask.ca}

\author{Banani Roy}
\orcid{0000-0003-1247-7781}
\affiliation{%
  \institution{University of Saskatchewan}
  \department{Computer Science}
  \city{Saskatoon}
  \country{Canada}
}
\email{banani.roy@usask.ca}

\author{Kevin A. Schneider}
\orcid{0000-0003-1113-1754}
\affiliation{%
  \institution{University of Saskatchewan}
  \department{Computer Science}
  \city{Saskatoon}
  \country{Canada}
}
\email{kevin.schneider@usask.ca}

\author{Chanchal K. Roy}
\orcid{0000-0003-0519-6164}
\affiliation{%
  \institution{University of Saskatchewan}
  \department{Computer Science}
  \city{Saskatoon}
  \country{Canada}
}
\email{chanchal.roy@usask.ca}

\begin{abstract}
The growing diversity of code clone types, from syntactic copies to cross-language semantic clones to AI-generated duplicates, has created a fragmentation crisis in clone detection. Current deep learning detectors are domain specialists that degrade significantly outside their training distribution. In our evaluation, F1 drops exceed 70\% across domains. Deploying multiple specialized models is impractical, yet training a single cross-domain detector requires simultaneous access to all training data. To address this, we investigate model merging, a family of post-hoc techniques that operate solely on trained checkpoints. We systematically evaluate parameter merging with five task-vector methods (Task Arithmetic, TIES, DARE-TIES, WUDI, PCB), architecture merging via greedy layer stitching, and cross-tokenizer alignment. Our study spans four pre-trained code models, three benchmarks, and twelve merging configurations. Same-base TIES merging creates effective cross-domain detectors, validated across two model families and three random seeds. It reaches 0.865 combined F1 on UniXcoder, 93\% of multi-task training performance without any training data at the merging step. WUDI achieves the highest in-distribution combined F1 at 0.899, but TIES generalizes better to unseen AI-generated clones. We therefore recommend TIES for the practical recipe. Cross-base merging yields only marginal and high-variance gains across all five methods. This indicates that task vector compatibility through a shared pre-trained base is the binding factor for effective merging in our setting. Merged encoder-based detectors also outperform zero-shot instruction-tuned code LLMs on GPTCloneBench (0.609 vs.\ 0.454 F1 for the strongest LLM baseline) at substantially lower per-pair inference cost. They also generalize up to 4$\times$ better than multi-task training to unseen AI-generated clones, suggesting a trade-off between in-domain performance and OOD robustness. This work provides one of the first systematic empirical studies of model merging for software engineering and a practical recipe for building cross-domain code clone detectors.
\end{abstract}




\maketitle

\section{Introduction}
\label{sec:introduction}
Code clone detection is fundamental to software maintenance, quality assurance, and evolution~\cite{roy2007survey, rattan2013software, roy2025jit}. The field has made remarkable progress over the past two decades, moving from token-based and tree-based matching~\cite{kamiya2002ccfinder, baxter1998clone, jiang2007deckard}, through learned representations~\cite{white2016deep, zhang2019novel, wei2017supervised}, to pre-trained code models that achieve strong results on established benchmarks~\cite{feng2020codebert, guo2021graphcodebert, guo2022unixcoder, lu2021codexglue}. This progress has come with an underappreciated cost. Each new detection capability, whether for cross-language clones~\cite{nafi2019clcdsa, perez2019cross} or AI-generated semantic clones~\cite{alam2023gptclonebench, roy2023unveiling}, produces a new specialist model that works well in its own domain but fails outside it. Teams must therefore deploy and maintain multiple specialized models, each limited to a single clone type or language pair.

The consequence is a growing \textit{out-of-distribution} (OOD) generalization problem~\cite{wang2022generalizing, hendrycks2020pretrained}. UniXcoder~\cite{guo2022unixcoder} fine-tuned on BigCloneBench~\cite{svajlenko2014towards} achieves 0.940 F1 on same-language clones but collapses to 0.269 on CLCDSA~\cite{nafi2019clcdsa} cross-language detection, a 71\% drop. The reverse is equally severe. This is a well-documented limitation of fine-tuned models under distribution shift~\cite{hendrycks2020pretrained, nguyen2024deep}, and recent work confirms similar degradation for clone detectors on unseen functionalities~\cite{kitsios2025detecting, li2023zc3} and highlights its implications for software engineering (SE)~\cite{berend2020cats, nguyen2024deep}. What is notable is that, despite the practical urgency, no prior work has investigated how this cross-domain fragmentation can be addressed for clone detection without retraining.

The obvious remedy of training a single model on all available data has its own limitations~\cite{vandenhende2021multi}. Multi-task learning requires simultaneous access to every training corpus, full retraining whenever a new domain appears, and careful loss balancing to avoid negative transfer~\cite{crawshaw2020multi}. These requirements are impractical in many SE settings where training data may be proprietary, distributed across teams, or simply too expensive to combine~\cite{yang2024federated}. This raises a question. \textit{How can we unify existing specialist detectors into a single cross-domain model without retraining and without access to any training data at the merging step?}

Model merging~\cite{ilharco2022editing, yadav2023ties, yu2024language} offers a principled path forward. These techniques combine trained models by operating directly on their checkpoints, requiring no training data, no gradient computation, and no retraining for the merging step itself. Merging completes in under five minutes on a single CPU in our experiments. Model merging has shown promise in NLP~\cite{yadav2023ties, zhou2024hm3} and computer vision~\cite{wortsman2022model, ilharco2022editing}, but its applicability to software engineering tasks remains unexplored. The merging techniques themselves are established, yet no prior work has investigated their applicability to software engineering or identified the conditions under which they succeed or fail for code-related tasks.

We present the first systematic empirical investigation of model merging for code clone detection, spanning parameter merging, architecture merging, and cross-tokenizer alignment across multiple model families. Our study evaluates parameter merging with five task-vector methods (Task Arithmetic~\cite{ilharco2022editing}, TIES~\cite{yadav2023ties}, DARE-TIES~\cite{yu2024language}, WUDI~\cite{cheng2025wudi}, and PCB~\cite{du2024parameter}) and simple averaging~\cite{wortsman2022model}, architecture merging via greedy layer stitching, and cross-tokenizer representation alignment. This spans four pre-trained code models, three benchmarks, and twelve merging configurations. Same-base TIES merging produces a cross-domain detector that achieves 0.890 $\pm$ 0.010 F1 on same-language clones and 0.839 $\pm$ 0.027 on cross-language clones simultaneously across three seeds, a 25\% combined improvement over the best individually fine-tuned model.

Same-base WUDI achieves the highest in-distribution combined F1 in our experiments (0.899 $\pm$ 0.025), but TIES generalizes better to unseen AI-generated clones, which informs our recommendation of TIES for the practical recipe. Cross-base merging yields only marginal, high-variance gains, which we trace to destructive interference between near-orthogonal task vectors.

Merging also holds up against strong alternatives. Multi-task training on the combined data wins in-domain (0.927 vs.\ 0.865 combined F1) but collapses on unseen AI-generated clones (0.151 F1), where merging generalizes far better. This suggests joint training overfits the combined distribution while merging retains complementary specialist knowledge. Two zero-shot instruction-tuned code LLMs, Qwen2.5-Coder-7B-Instruct and DeepSeek-Coder-6.7B-Instruct, also fall short of the merged detector, which runs orders of magnitude faster per pair. Finally, architecture-level layer stitching is interpretable: cross-language layers dominate early transformer positions and same-language layers dominate late ones. A learned cross-tokenizer alignment fails to bridge incompatible vocabularies, pointing to a tokenizer limitation that linear alignment cannot resolve. We validate these findings across two model families (UniXcoder and CodeBERT).

\noindent\textit{\textbf{RQ1.} How well do individual code clone detectors generalize across detection domains?} 
We quantify the cross-domain gap to establish the severity of the OOD problem.

\noindent\textit{\textbf{RQ2.} What conditions determine successful parameter-level merging for cross-domain code clone detection?} 
We compare same-base and cross-base merging across five task-vector methods and benchmark against multi-task training and two zero-shot LLM baselines.

\noindent\textit{\textbf{RQ3.} What representations are captured by different layers across domains, and how can layer stitching exploit them?} 
We explore whether layer-level selection captures domain-specific roles that uniform merging cannot.

\noindent\textit{\textbf{RQ4.} Do merged models generalize to unseen clone types, such as LLM-generated semantic clones?} 
We test zero-shot robustness on AI-generated semantic clones unseen during training.
\newline We make the following contributions.
\begin{enumerate*}
    \item We provide the first systematic empirical study of post-hoc model merging for cross-domain code clone detection, and identify shared-base task-vector compatibility as the binding factor for effective merging without any training data at the merging step.
    \item We give geometric and statistical evidence for why cross-base merging fails and same-base merging succeeds, and show this pattern holds across all five task-vector methods we evaluate, establishing that the compatibility boundary is method-independent rather than an artifact of any single algorithm.
    \item We show that architecture-level layer stitching reveals interpretable domain-specific patterns, and identify a tokenizer barrier that linear cross-tokenizer alignment cannot overcome.
    \item We distill a practical, training-data-free recipe: fine-tune one shared base per domain and merge with TIES at the parameter level. We discuss its scope and limitations in Section~\ref{sec:discussion}.
\end{enumerate*}
\begin{figure}[H]
\centering
\includegraphics[width=\columnwidth]{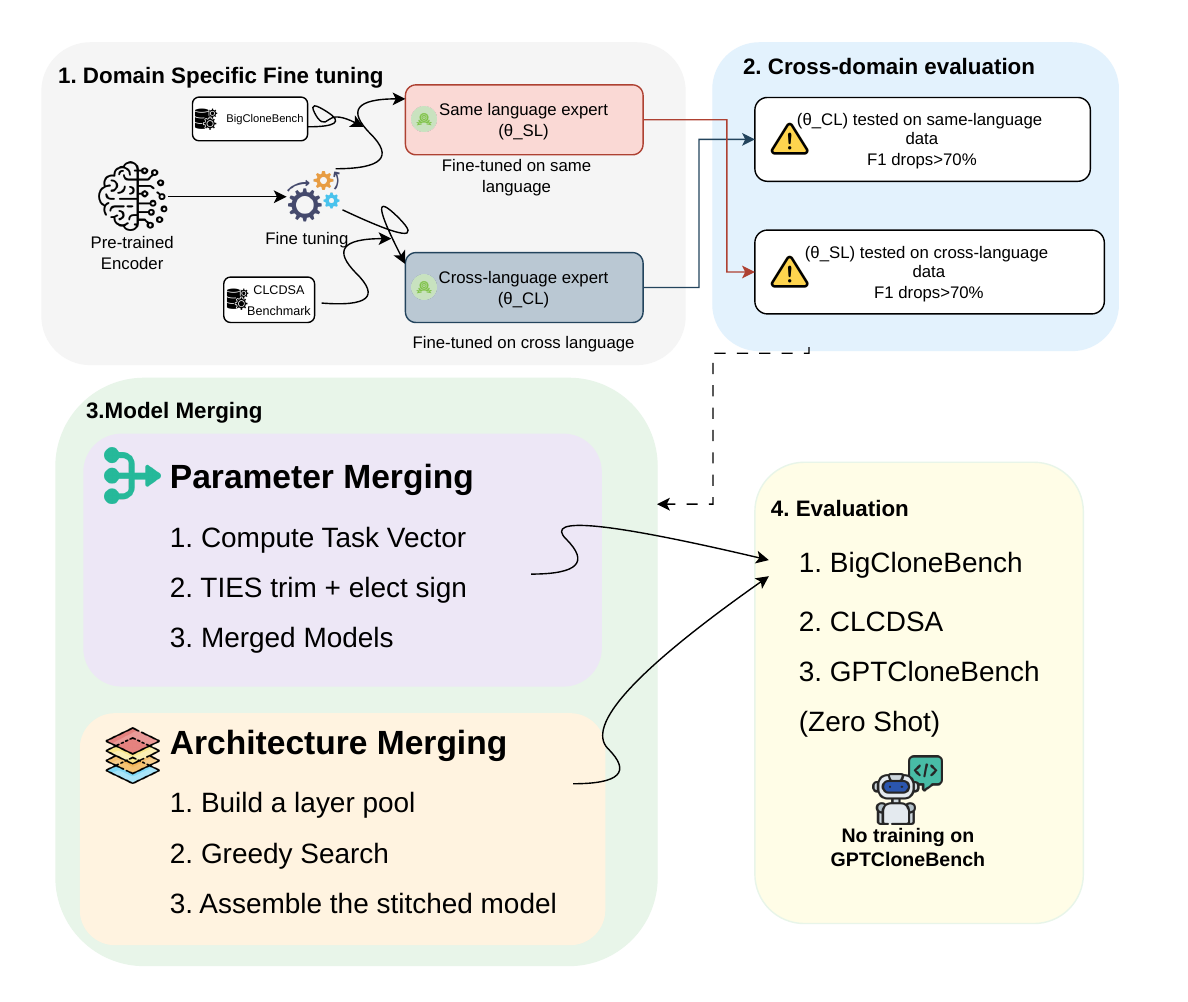}
\caption{Overview of our four-phase approach.}
\label{fig:methodology}
\end{figure}

\section{Background}\label{sec:background}

\textbf{Code Clones.} Code clones are similar code fragments that arise through copy-paste, independent development, or automated generation~\cite{roy2007survey}. The standard taxonomy~\cite{roy2007survey, rattan2013software} ranges from exact copies (Type-1) through renamed (Type-2) and modified (Type-3) fragments to functionally equivalent but syntactically different code (Type-4). Two specialized forms of Type-4 clones are increasingly important. \textit{Cross-language clones} are functionally equivalent implementations across different languages~\cite{nafi2019clcdsa}, and \textit{AI-generated semantic clones} are code produced by large language models with distinct syntactic patterns~\cite{alam2025classical, alam2023gptclonebench, roy2023unveiling}. Each requires its own detector and training data, creating the cross-domain fragmentation problem we address. Our focus is on generalization across these detection domains, which is distinct from Type-1 through Type-4 generalization within a single domain.



\noindent\textbf{Model Merging.} Model merging combines multiple trained models into a single model without additional training~\cite{ilharco2022editing}. Its key abstraction is the \textit{task vector}, the difference between a fine-tuned model's parameters and its pre-trained base. Task vectors encode domain-specific knowledge and can be arithmetically combined to transfer it across tasks~\cite{ilharco2022editing}. Several methods refine naive task-vector addition. TIES~\cite{yadav2023ties} resolves sign conflicts through trimming and sign election, DARE~\cite{yu2024language} drops and rescales elements to reduce redundancy, WUDI~\cite{cheng2025wudi} minimizes per-layer cross-task interference without training data, and PCB~\cite{du2024parameter} balances parameter competition before rescaling. These methods succeed in NLP and computer vision, but their applicability to software engineering remains largely unexplored.


\section{Approach}
\label{sec:approach}

Figure~\ref{fig:methodology} presents an overview of our approach. We investigate model merging for code clone detection through a four-phase pipeline: (1) domain-specific fine-tuning of specialist models, (2) cross-domain evaluation to quantify the generalization gap, (3) model merging through parameter-level and architecture-level strategies, and (4) evaluation on both in-domain and unseen benchmarks.

\subsection{Domain-Specific Fine-Tuning}
\label{sec:finetuning}
We formulate clone detection as a binary classification task over code pairs. Given a pair $(x_1, x_2)$, we concatenate the two code snippets as \texttt{[CLS] $x_1$ [SEP] $x_2$ [SEP]} and pass them through the encoder. The \texttt{[CLS]} representation is fed through a dropout layer and a linear classification head ($768 \rightarrow 2$) trained using cross-entropy loss. Given a pre-trained encoder $\theta_{\text{base}}$, we fine-tune on a target dataset $\mathcal{D}_k$ to obtain a specialist checkpoint $\theta_k$ optimized for a single detection domain. This yields $K$ specialist models. Each one achieves high performance on its respective domain but degrades severely on out-of-distribution inputs. Before merging, we evaluate each specialist on all benchmarks to quantify the cross-domain gap and identify complementary models.

\subsection{Parameter-Level Merging}
\label{sec:param_merging}

Parameter-level merging combines the weights of multiple specialist models into a single unified model through arithmetic operations on their parameters, without requiring any training data or gradient computation at the merging step.

\subsubsection{Task Vectors}
Given a fine-tuned model $\theta_k$ and its corresponding pre-trained base $\theta_{\text{base}}$, the \textit{task vector}~\cite{ilharco2022editing} is defined as
\vspace{-0.2cm}
\begin{equation}
\delta_k = \theta_k - \theta_{\text{base}}
\label{eq:task_vector}
\end{equation}
The task vector $\delta_k$ encodes the knowledge gained during fine-tuning on domain $k$. A merged model is constructed by adding a weighted combination of task vectors back to the base model,
\begin{equation}
\theta_{\text{merged}} = \theta_{\text{base}} + 
\sum_{k=1}^{K} \lambda_k \cdot \delta_k
\label{eq:merge}
\end{equation}
where $\lambda_k$ are merging weights that control the relative contribution of each specialist. In our experiments, we set $\lambda_k = 1/K$ for all specialists unless otherwise specified.

\subsubsection{TIES Merging}
Naively summing task vectors leads to interference when parameters from different specialists conflict in sign or magnitude. TIES (TrIm, Elect Sign, and Merge)~\cite{yadav2023ties} addresses this through three steps: (1)~\textit{Trim}: zero out task vector values below a percentile threshold, retaining only the most significant parameter changes. (2)~\textit{Elect sign}: for each parameter position, compute a weighted vote across all task vectors and elect the majority sign direction. (3)~\textit{Merge}: sum only the task vector values whose signs agree with the elected direction, discarding conflicting updates. The merged model becomes
\begin{equation}
\theta_{\text{merged}} = \theta_{\text{base}} + 
\sum_{k=1}^{K} \lambda_k \cdot 
\text{TIES}(\delta_k, \{\delta_j\}_{j=1}^{K})
\label{eq:ties}
\end{equation}

\subsubsection{DARE Dropping}
DARE (Drop And REscale)~\cite{yu2024language} extends task vector merging by randomly dropping a fraction $p$ of each task vector's elements and rescaling the surviving values by $1/(1-p)$,
\begin{equation}
\tilde{\delta}_k = \frac{1}{1-p} \cdot 
(m \odot \delta_k), \quad 
m_i \sim \text{Bernoulli}(1-p)
\label{eq:dare}
\end{equation}
where $\odot$ denotes element-wise multiplication. DARE can be combined with TIES to form DARE-TIES, where dropping is applied before the trim-elect-merge steps.

\subsubsection{WUDI Merging}
\label{sec:wudi}
WUDI~\cite{cheng2025wudi} casts merging as a per-layer optimization rather than a heuristic conflict-resolution rule. For each linear layer, it searches for the merged task vector that minimizes cross-task interference, without training data or rescaling coefficients. The objective for a layer with task vectors $\{\delta_k\}_{k=1}^{K}$ is
\begin{equation}
\mathcal{L}_{\text{WUDI}}(\delta) = 
\sum_{k=1}^{K} \frac{\|(\delta - \delta_k)\, \delta_k^{\top}\|_F^2}{\|\delta_k\|_F^2},
\label{eq:wudi}
\end{equation}
where $\|\cdot\|_F$ is the Frobenius norm, and the merged model is $\theta_{\text{merged}} = \theta_{\text{base}} + \arg\min_{\delta}\mathcal{L}_{\text{WUDI}}(\delta)$. We initialize $\delta$ as the task-arithmetic sum and optimize with Adam ($1\!\times\!10^{-5}$, 300 steps) per linear layer, applying WUDI to the encoder and classifier-head weight matrices and averaging non-linear parameters (layer normalization, embeddings, biases) as in TIES.

\subsubsection{PCB Merging}
\label{sec:pcb}
PCB~\cite{du2024parameter} is a training-free task-vector method that balances parameter competition within and across task vectors, dropping low-competition parameters before rescaling. We include it as a fifth method to test whether the same-base compatibility boundary holds beyond sign election (TIES) and interference minimization (WUDI).

\subsubsection{Same-Base Hypothesis}
\label{sec:same_base}

We hypothesize that effective parameter merging requires all specialist models to share the same pre-trained base $\theta_{\text{base}}$. When specialists are fine-tuned from different base models (e.g., CodeBERT vs.\ UniXcoder), their task vectors reside in different parameter spaces, and arithmetic combination may produce destructive interference rather than constructive merging. Whether the shared-base principle established in NLP and vision~\cite{ilharco2022editing, yadav2023ties, cheng2025wudi} holds for code clone detection, where models differ in pre-training objectives and tokenizers, remains an open empirical question. We validate this hypothesis across all five task-vector methods (Task Arithmetic, TIES, DARE-TIES, WUDI, and PCB) in Section~\ref{sec:results}.

\subsection{Architecture-Level Merging}
\label{sec:arch_merging}

While parameter merging operates uniformly across all layers, different layers of a transformer may benefit from different specialists. Architecture-level merging addresses this by selecting, for each layer position, which specialist's layer to use in the final composite model.

\subsubsection{Layer Pool Construction}
Given $K$ specialist models each with $L$ transformer layers, we construct a \textit{layer pool} 
$\mathcal{P} = \{(k, l) \mid k \in [1, K], 
l \in [0, L{-}1]\}$ containing all candidate layers. For two specialists, this yields $2L$ candidate layers for $L$ positions.

\subsubsection{Greedy Layer Selection}
We employ a greedy search to construct the optimal \textit{inference path} through the layer pool. Starting from the best-performing individual specialist as the initial model, we iterate over each layer position $l = 0, 1, \ldots, L{-}1$ and evaluate whether swapping in an alternative specialist's layer improves a combined evaluation objective,
\begin{equation}
\mathcal{F} = \frac{1}{2}\left(
\text{F1}_{\text{SL}} + \text{F1}_{\text{CL}}\right)
\label{eq:combined_f1}
\end{equation}
We compute $\mathcal{F}$ on held-out validation sets for each domain. Greedy search requires only $K \times L$ evaluations (24 in our setting) compared to $K^L$ for exhaustive enumeration, while producing strong results in practice. At each position, the swap that yields the highest $\mathcal{F}$ is retained. If no swap improves over the current configuration, the original layer is kept. This produces a composite model where each layer is drawn from whichever specialist contributes most to the combined objective at that position.

\subsubsection{Tokenizer Constraint}
Architecture-level merging requires that all candidate models share the same tokenizer. Different tokenizers produce different token ID sequences for the same source code, making the embedding layers incompatible. We investigate whether this constraint can be relaxed through representation alignment (Section~\ref{sec:alignment}).

\subsection{Cross-Tokenizer Representation Alignment}
\label{sec:alignment}

To enable merging across models with different tokenizers, we explore learning a linear projection that maps one model's hidden representations into the other's space.

Given a source model $\theta_S$ (e.g., UniXcoder, vocabulary size 51,416) and a target model $\theta_T$ (e.g., CodeBERT, vocabulary size 50,265), we collect paired hidden states by feeding the same raw code samples through both models using their respective tokenizers. We use a separate subset of training data, disjoint from downstream evaluation, to avoid leakage. For each transformer layer $l$, we train a linear projection $W_l \in \mathbb{R}^{d \times d}$ that minimizes
\begin{equation}
\mathcal{L}_{\text{align}} = 
\|W_l \cdot h_S^{(l)} - h_T^{(l)}\|_2^2 + 
\alpha \cdot \mathcal{L}_{\text{CORAL}}
\label{eq:align}
\end{equation}
where $h_S^{(l)}$ and $h_T^{(l)}$ are the mean-pooled hidden states at layer $l$ from the source and target models respectively. We use mean pooling to obtain stable sequence-level representations across variable-length inputs. $\mathcal{L}_{\text{CORAL}}$ is the Deep CORAL loss~\cite{sun2016deep} that aligns the covariance structures of the two representation spaces.

Once trained, the projection $W_l$ is inserted as a forward hook after layer $l$ whenever a source model layer is selected by the greedy search, transforming the output into the target model's representation space before passing to the next layer.

\section{Experimental Setup}
\label{sec:setup}

\textbf{Datasets.}\label{sec:datasets} Table~\ref{tab:datasets} summarizes the three benchmarks used in our evaluation.

\begin{table}[t]
\centering
\caption{Summary of evaluation benchmarks.}
\label{tab:datasets}
\begin{tabular}{llrrr}
\toprule
\textbf{Dataset} & \textbf{Domain} & \textbf{Train} & 
\textbf{Test} & \textbf{Lang.} \\
\midrule
BCB & Same-lang & 901,028 & 415,416 & Java \\
CLCDSA data & Cross-lang & 484,676 & 111,032 & Java, Py \\
GPTCloneBench & AI-Clone & --- & 10,000 & Java \\
\bottomrule
\end{tabular}
\end{table}

\textit{BigCloneBench (BCB)}~\cite{svajlenko2014bigclonebench} is the most widely used benchmark for same-language code clone detection. We use the CodeXGLUE split~\cite{lu2021codexglue} in Java. The dataset exhibits class imbalance with approximately 14\% positive clone pairs, which we address using weighted random sampling proportional to inverse class frequency.

\textit{CLCDSA data}~\cite{nafi2019clcdsa} is a cross-language clone detection benchmark containing solutions in multiple programming languages to competitive programming problems from AtCoder and CodeJam. We use the Java$\leftrightarrow$Python subset throughout our experiments, in which each pair consists of one Java source file and one Python source file. We apply a problem-level split (70/15/15) to ensure no problem overlap between splits, and negative pairs are constructed by sampling Python solutions from different problems. The resulting subset is approximately balanced with 50\% positive pairs by construction.

\textit{GPTCloneBench}~\cite{alam2023gptclonebench} contains GPT-generated semantic clones built upon SemanticCloneBench~\cite{al2020semanticclonebench}. We construct 10,000 balanced evaluation pairs by randomly sampling from the Java subset, using 5,000 true clones from intra-prompt pairs and 5,000 non-clones from cross-functionality pairs, with a fixed random seed for reproducibility. \textit{No model is trained on this benchmark.} It serves exclusively for zero-shot generalization evaluation.

\noindent\textbf{Subject Models.}\label{sec:models} We select three pre-trained code models based on the RoBERTa encoder architecture, each with 12 transformer layers and 768-dimensional hidden representations.

\textit{CodeBERT}~\cite{feng2020codebert} is pre-trained on code-natural language pairs in six programming languages using masked language modeling and replaced token detection. \textit{GraphCodeBERT}~\cite{guo2021graphcodebert} extends CodeBERT by incorporating data flow graphs during pre-training. Both share the same tokenizer with vocabulary size 50,265. \textit{UniXcoder}~\cite{guo2022unixcoder} is a unified cross-modal model that uses AST-based pre-training objectives for code understanding and generation, with a distinct tokenizer of vocabulary size 51,416.

We fine-tune CodeBERT and GraphCodeBERT on BigCloneBench, and fine-tune UniXcoder on both BigCloneBench and the CLCDSA Java$\leftrightarrow$Python subset, producing four specialist models (Table~\ref{tab:specialists}). We focus on UniXcoder for cross-language fine-tuning because it incorporates explicit cross-modal pre-training objectives~\cite{guo2022unixcoder}, whereas CodeBERT and GraphCodeBERT use NL-PL pair pre-training. This design enables our central comparison between same-base merging (two UniXcoder variants) and cross-base merging (CodeBERT, GraphCodeBERT, and UniXcoder).

\begin{table}[t]
\centering
\caption{Specialist models and their validation F1 scores.}
\label{tab:specialists}
\begin{tabular}{llr}
\toprule
\textbf{Model} & \textbf{Training Data} & 
\textbf{Valid. F1} \\
\midrule
CodeBERT & BCB & 0.895 \\
GraphCodeBERT & BCB & 0.905 \\
UniXcoder-BCB & BCB & 0.931 \\
UniXcoder-CLCDSA & CLCDSA data & 0.940 \\
\bottomrule
\end{tabular}
\end{table}

\noindent\textbf{Implementation Details.}
\label{sec:implementation}
All specialist models are fine-tuned with identical hyperparameters. We use the AdamW optimizer with learning rate $2 \times 10^{-5}$, batch size 32, maximum sequence length 512, 10\% linear warmup, and early stopping with patience of 3 epochs (up to 10 epochs maximum). Training uses mixed precision (bfloat16) on NVIDIA A100 80GB PCIe GPUs. For the multi-task baseline, we fine-tune UniXcoder on the combined BCB and CLCDSA data training sets (1.38M pairs) using identical hyperparameters. For the CodeBERT validation experiment (Section~\ref{sec:codebert_validation}), we fine-tune CodeBERT on CLCDSA data using the same settings. To assess robustness, all specialist models involved in same-base merging are trained across three random seeds (42, 2, 3). We report mean $\pm$ standard deviation for merged model performance.

For parameter merging, we evaluate multiple configurations as detailed in Section~\ref{sec:results}. We use DARE-TIES (drop rate $p = 0.3$), TIES-only ($p = 0$), and simple averaging (no trimming), with merging weights $\lambda_k = 1/K$. For WUDI merging~\cite{cheng2025wudi}, we follow the official RoBERTa implementation with the Adam optimizer (learning rate $1 \times 10^{-5}$, 300 steps per linear layer) and the merged delta initialized as the task-arithmetic sum. Non-linear parameters (LayerNorm, embeddings, biases) use weighted averaging with $\lambda_k = 1/K$. For PCB merging~\cite{du2024parameter}, we follow the official implementation with drop rate 0.1 and intra-balancing scale 1.0, applying PCB to the linear weight matrices and averaging non-linear parameters as above. For architecture merging, the greedy layer search evaluates each candidate swap on 1,000 held-out validation samples per dataset. For cross-tokenizer alignment, we train 12 linear projections ($768 \times 768$) on 5,000 code samples from the BigCloneBench training set (disjoint from evaluation data) using MSE + CORAL loss ($\alpha = 0.1$) for 20 epochs.

All models are evaluated using identical procedures with the same test splits, the same classification threshold (argmax over logits), and the same metric computation pipeline. All model selection and merging decisions are performed on validation sets, and final performance is reported on held-out test sets throughout Section~\ref{sec:results}.

\noindent\textbf{LLM Baseline.}
\label{sec:llm_baseline}
To compare our merged encoder-based detector against representative instruction-tuned code LLMs, we evaluate two models, Qwen2.5-Coder-7B-Instruct~\cite{hui2024qwen25coder} and DeepSeek-Coder-6.7B-Instruct~\cite{guo2024deepseek}, in a zero-shot setting on GPTCloneBench using the official model releases without any task-specific fine-tuning. For each evaluation pair $(x_1, x_2)$, we apply the model's chat template with a fixed system prompt (``You are an expert software engineer specializing in code clone detection.'') and a user prompt that presents the two snippets and asks the model to answer with a single word, ``Yes'' or ``No'', indicating whether the snippets are semantic clones. Both models receive an identical prompt, decoding configuration, and evaluation set; only their respective instruction formats differ. We decode greedily (\texttt{do\_sample=False}) with a maximum of 32 new tokens and parse the Yes/No verdict from the generated response. Outputs that do not conform to the requested one-word format (0\% for Qwen, 13.6\% for DeepSeek) are scored as non-clone. If the combined token length exceeds the context budget, we truncate the two snippets proportionally to their lengths (0 of 10,000 pairs required truncation in our runs). Inference is data-parallel across NVIDIA A100 80GB PCIe GPUs in bfloat16, on the same 10,000 GPTCloneBench pairs and seed used throughout the paper. The full prompt template, per-pair predictions, and inference scripts are in the replication package (Section~\ref{sec:das}).

\noindent\textbf{Evaluation Metrics.}
\label{sec:metrics}
We report F1 score, precision, and recall for binary clone detection on each benchmark's full test set. To measure cross-domain effectiveness, we define the combined F1 as
\begin{equation}
\text{F1}_{\text{comb}} = \frac{1}{2}\left(
\text{F1}_{\text{SL}} + \text{F1}_{\text{CL}}\right)
\label{eq:combined_metric}
\end{equation}
where $\text{F1}_{\text{SL}}$ and $\text{F1}_{\text{CL}}$ are F1 scores on BigCloneBench and CLCDSA data test sets respectively. We use equal weighting in the absence of a domain-specific preference.

\section{Results}
\label{sec:results}

\textbf{RQ1. Cross-Domain Generalization Gap}\label{sec:rq1}

\begin{table}[t]
\centering
\caption{Cross-domain evaluation of specialist models. Bold = in-domain.}
\label{tab:cross_eval}
\begin{tabular}{lccc}
\toprule
\textbf{Model} & \textbf{BCB F1} & \textbf{CLCDSA F1} & 
\textbf{F1\textsubscript{comb}} \\
\midrule
CodeBERT & \textbf{0.894} & 0.464 & 0.679 \\
GraphCodeBERT & \textbf{0.891} & 0.495 & 0.693 \\
UniXcoder-BCB & \textbf{0.940} & 0.269 & 0.605 \\
UniXcoder-CLCDSA & 0.259 & \textbf{0.924} & 0.592 \\
\bottomrule
\end{tabular}
\end{table}

\noindent Table~\ref{tab:cross_eval} presents the cross-domain evaluation results. All four specialist models achieve strong in-domain performance with F1 $\geq$ 0.891, confirming that fine-tuning produces effective domain-specific detectors. Cross-domain performance, however, degrades severely. UniXcoder-BCB achieves 0.940 F1 on BigCloneBench but only 0.269 on CLCDSA, a 71\% drop. Conversely, UniXcoder-CLCDSA achieves 0.924 on CLCDSA but collapses to 0.259 on BigCloneBench, a 72\% drop. CodeBERT and GraphCodeBERT exhibit a similar pattern with 44--48\% drops on CLCDSA.

This degradation confirms the out-of-distribution challenge described in Section~\ref{sec:introduction}. Each model's learned decision boundary is specific to its training distribution and does not transfer across clone detection domains. The best combined F1 among individual models is only 0.693 (GraphCodeBERT), indicating that no single specialist adequately covers both domains despite achieving up to 0.940 F1 on its home benchmark.

\begin{tcolorbox}[before skip=5pt, after skip=5pt, colback=gray!5, colframe=gray!50, 
boxrule=0.5pt, left=4pt, right=4pt, top=3pt, bottom=3pt]\textbf{Finding 1.} Individual clone detectors suffer 44--72\% F1 drops when evaluated outside their training domain, confirming the OOD generalization failure. No individual model achieves a combined F1 above 0.693.
\end{tcolorbox}

\noindent\textbf{RQ2. Parameter-Level Merging}\label{sec:rq2}

\begin{table*}[!htbp]
\centering
\caption{Parameter merging results. Three-seed mean ± std for same-base TIES/WUDI/PCB and cross-base WUDI/PCB.}
\label{tab:param_merge}
\begin{tabular}{llcccc}
\toprule
\textbf{Category} & \textbf{Configuration} & 
\textbf{Drop $p$} & \textbf{BCB F1} & 
\textbf{CLCDSA F1} & \textbf{F1\textsubscript{comb}} \\
\midrule
\multirow{6}{*}{Cross-base} 
& DARE-TIES (CB/GCB/UX, $\lambda$=0.4/0.3/0.3) 
& 0.3 & 0.864 & 0.332 & 0.598 \\
& TIES-only (CB/GCB/UX, $\lambda$=0.4/0.3/0.3) 
& 0.0 & 0.841 & 0.566 & 0.703 \\
& Equal weights (CB/GCB/UX, $\lambda$=0.33 each) 
& 0.3 & 0.502 & 0.017 & 0.260 \\
& UniX-heavy (CB/GCB/UX, $\lambda$=0.2/0.2/0.6) 
& 0.3 & 0.241 & 0.619 & 0.430 \\
& WUDI (UX-BCB + CB-CLCDSA, $\lambda$=0.5/0.5) 
& --- & .939\tiny{$\pm$.001} & .372\tiny{$\pm$.148} & .656\tiny{$\pm$.074} \\
& PCB (UX-BCB + CB-CLCDSA, $\lambda$=0.5/0.5) 
& --- & .935\tiny{$\pm$.004} & .542\tiny{$\pm$.122} & .739\tiny{$\pm$.060} \\
\midrule
\multirow{4}{*}{Same-base} 
& Simple averaging (UX+UX, $\lambda$=0.5/0.5) 
& 0.0 & 0.893 & 0.801 & 0.847 \\
& TIES (UX+UX, $\lambda$=0.5/0.5) 
& 0.0 & .890\tiny{$\pm$.010} & .839\tiny{$\pm$.027} & .865\tiny{$\pm$.012} \\
& PCB (UX+UX, $\lambda$=0.5/0.5) 
& --- & .910\tiny{$\pm$.030} & .854\tiny{$\pm$.040} & .882\tiny{$\pm$.009} \\
& \textbf{WUDI (UX+UX, $\lambda$=0.5/0.5)} 
& --- & \textbf{.935}\tiny{$\pm$.004} & \textbf{.863}\tiny{$\pm$.052} & 
\textbf{.899}\tiny{$\pm$.025} \\
\bottomrule
\end{tabular}
\end{table*}

Table~\ref{tab:param_merge} presents parameter merging results across ten configurations.

\textbf{Cross-base merging is consistently weaker and less stable than same-base merging.} The best cross-base configurations, PCB (0.739 $\pm$ 0.060) and TIES-only (0.703), exceed the best individual model (0.693) only modestly and fall well short of every same-base merge, while weaker weightings collapse entirely (down to 0.260). The two three-seed methods, WUDI (0.656 $\pm$ 0.074) and PCB (0.739 $\pm$ 0.060), share an asymmetric, unstable profile: BCB F1 stays high as the UniXcoder-BCB specialist dominates its own parameter space, while CLCDSA F1 shows a lower mean and far higher variance (up to $\pm$0.148) than its same-base counterpart. This limitation is not specific to TIES or DARE. It persists under WUDI's interference-minimization and PCB's competition-balancing objectives, indicating the compatibility constraint is method-independent.

\textbf{Same-base merging is highly effective.} When both specialists share the same pre-trained base (UniXcoder), even simple averaging achieves 0.847 combined F1, already surpassing all cross-base configurations. TIES further improves this to 0.865 $\pm$ 0.012 across three seeds and PCB to 0.882 $\pm$ 0.009. WUDI, which directly optimizes the merged task vector per linear layer to minimize cross-task interference, achieves the highest in-domain combined F1 at 0.899 $\pm$ 0.025, retaining 96--99\% of each specialist's in-domain performance without any training data at the merging step.

The gap between TIES and simple averaging indicates that sign conflict resolution contributes meaningfully even when task vectors share the same base. WUDI's further gain shows that direct interference-minimization extracts additional in-distribution signal compared to sign-election heuristics, with PCB falling between. However, WUDI's OOD behavior differs from TIES (Section~\ref{sec:rq4}), which informs our practical recommendation. Section~\ref{sec:why_same_base} analyzes the geometric properties of same-base and cross-base task vectors.

\textbf{Comparison with multi-task training.} To assess whether the merging step can compete with training on all data simultaneously, we fine-tune UniXcoder on the combined BCB and CLCDSA data (1.38M pairs). The multi-task model achieves higher in-domain performance (combined 0.927), as expected. Same-base TIES achieves 93\% of this and same-base WUDI 97\%, without any training data at the merging step. However, the multi-task model underperforms substantially on GPTCloneBench (0.151 F1), while same-base TIES-merged models generalize far better (0.450--0.609 mean F1 across three seeds). Training on all domains simultaneously leads to overfitting the combined distribution, whereas merging retains independently optimized specialist knowledge.

\begin{tcolorbox}[before skip=5pt, after skip=5pt, colback=gray!5, colframe=gray!50, boxrule=0.5pt, left=4pt, right=4pt, top=3pt, bottom=3pt]
\textbf{Finding 2.} Same-base merging is method-independent and consistently outperforms cross-base merging across all five task-vector methods. Same-base TIES reaches 93\% of multi-task performance without any training data and generalizes 4$\times$ better to unseen GPT-generated clones.
\end{tcolorbox}

\noindent\textbf{RQ3. Architecture-Level Merging}\label{sec:rq3}
\noindent Table~\ref{tab:arch_merge} compares architecture merging with parameter merging across three scenarios. The relative effectiveness depends strongly on model compatibility (same-base vs.\ cross-base).

\begin{table}[H]
\centering
\caption{Architecture merging compared to parameter merging.}
\label{tab:arch_merge}
\begin{tabular}{lccc}
\toprule
\textbf{Configuration} & \textbf{BCB} & 
\textbf{CLCDSA} & \textbf{F1\textsubscript{comb}} \\
\midrule
\multicolumn{4}{l}{\textit{Same-tokenizer, 
different base}} \\
CB+GCB (param.) & 0.841 & 0.566 & 0.703 \\
CB+GCB (arch.) & 0.856 & 0.620 & 0.738 \\
\midrule
\multicolumn{4}{l}{\textit{Same-base}} \\
UX+UX (arch.) & 0.842 & 0.743 & 0.793 \\
UX+UX (param.\ TIES) & 0.900 & 0.841 & 0.871 \\
UX+UX (param.\ WUDI) & \textbf{0.935} & \textbf{0.881} & \textbf{0.908} \\
\midrule
\multicolumn{4}{l}{\textit{Cross-tokenizer 
(with alignment)}} \\
CB+GCB+UX (arch.) & 0.856 & 0.620 & 0.738 \\
\bottomrule
\end{tabular}
\end{table}

\begin{figure}[!tbp]
\centering
\includegraphics[scale=0.8]{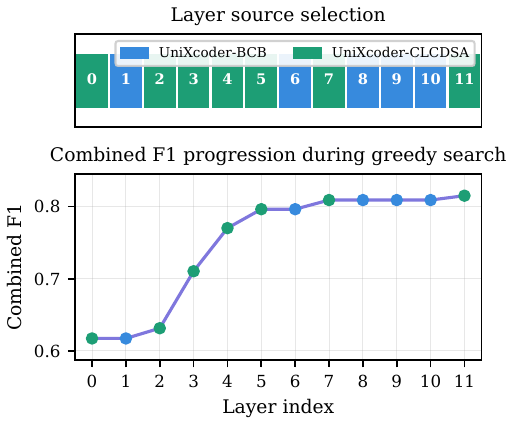}
\caption{Greedy layer selection for UniXcoder-BCB + UniXcoder-CLCDSA.}
\label{fig:layer_path}
\end{figure}

\textbf{Architecture merging outperforms parameter merging for cross-base models.} When merging CodeBERT and GraphCodeBERT (same tokenizer, different bases), greedy layer stitching (0.738) outperforms cross-base TIES (0.703) by 5\%. The greedy search identifies an interpretable pattern. CodeBERT layers are selected for early positions (layers 0--4, 7) while GraphCodeBERT layers are retained for later positions (layers 5--6, 8--11). This suggests that CodeBERT's early layers capture stronger token-level representations, while GraphCodeBERT's later layers leverage its data-flow pretraining for higher-level clone reasoning.

\textbf{For same-base models, parameter merging outperforms architecture merging.} Same-base UniXcoder parameter merges (TIES at 0.871, WUDI at 0.908) substantially outperform same-base architecture merging (0.793). When task vectors are compatible, uniform parameter combination across all layers is more effective than layer-level selection, which forces an all-or-nothing choice at each layer position. Parameter merging instead allows smooth interpolation across all parameters, enabling more effective integration of complementary knowledge.

\textbf{Cross-tokenizer alignment does not improve architecture merging.} Despite training linear projections with alignment loss below 0.005 for early layers (0--6), the greedy search never selected aligned UniXcoder layers over CodeBERT or GraphCodeBERT alternatives. Alignment loss increases sharply for later layers (0.090 for layer 10, 0.167 for layer 11), indicating that the representations diverge too substantially for linear projection to bridge. The tokenizer barrier may therefore reflect deeper differences in how models encode code semantics, not merely a vocabulary mismatch.

Figure~\ref{fig:layer_path} visualizes the layer selection path for the same-base UniXcoder merge. Cross-language (CLCDSA) layers dominate early and middle positions (layers 0, 2--5, 7, 11), while same-language (BCB) layers are retained at positions 1, 6, 8--10. This pattern suggests that cross-language fine-tuning produces stronger low-level code representations that capture language-agnostic features, while same-language fine-tuning specializes later layers for within-language clone discrimination. The validation-set combined F1 increases steadily from 0.617 (single specialist baseline) to 0.815 during the greedy search, with the largest gains at layers 3--5 where cross-language layers are introduced. The final held-out test performance of the resulting composite model is 0.793 (Table~\ref{tab:arch_merge}).

\begin{tcolorbox}[before skip=5pt, after skip=5pt, colback=gray!5, colframe=gray!50, 
boxrule=0.5pt, left=4pt, right=4pt, top=3pt, bottom=3pt]
\textbf{Finding 3.} Architecture merging outperforms parameter merging for cross-base models (+5\%) but underperforms for same-base models. Layer stitching reveals that cross-language layers dominate early positions while same-language layers dominate late positions. Linear cross-tokenizer alignment is insufficient.
\end{tcolorbox}

\noindent\textbf{RQ4. Generalization to GPT-Generated Clones}
\label{sec:rq4}
\noindent Table~\ref{tab:main_results} reports zero-shot GPTCloneBench~\cite{alam2023gptclonebench} F1 in its final column. All models were evaluated without any training or exposure to GPT-generated code. Despite the substantial domain shift from human-written to AI-generated clones, the same-base TIES merge and the architecture merge outperform every individual encoder specialist. The multi-task model trained on all available data achieves only 0.151 F1, substantially worse than all individual specialists and merged models, indicating severe overfitting to the combined training distribution.

Two merges generalize markedly better than any individual encoder. The same-base TIES merge of two CodeBERT specialists reaches the highest GPTCloneBench F1 at 0.609 across three seeds, a 35.3\% improvement over the best individual encoder (CodeBERT, 0.450), and the architecture merge of CodeBERT and GraphCodeBERT reaches 0.528 (+17.3\%). The same-base UniXcoder TIES merge is on par with the best encoder at 0.450.

\textbf{Comparison with zero-shot LLM baselines.} We evaluate two instruction-tuned code LLMs on the same 10,000 GPTCloneBench pairs (Section~\ref{sec:llm_baseline}). Qwen2.5-Coder-7B-Instruct achieves 0.454 F1 (precision 0.482, recall 0.429), comparable to the best individual encoder specialist but below the strongest merged detectors. DeepSeek-Coder-6.7B-Instruct is markedly more conservative under the identical prompt, predicting clone for only 510 of 10,000 pairs and reaching 0.120 F1 (precision 0.647, recall 0.066). Both LLMs fall below the same-base TIES merge of two CodeBERT specialists (0.609), which outperforms Qwen by 0.155 and DeepSeek by 0.489 F1 absolute. The cost gap is also large. Qwen inference required 628.9 seconds wall-clock for 10,000 pairs on three A100 GPUs, equivalent to 0.189 seconds per pair on a single-GPU basis, and DeepSeek is comparable. Encoder-based detectors process the same pairs in the order of milliseconds per pair on a single GPU. For repository-scale or CI-based clone detection workflows, a merged encoder detector is both more accurate and orders of magnitude cheaper than a zero-shot code LLM.

\textbf{WUDI and PCB show different generalization profiles.} While WUDI improves over TIES on in-distribution benchmarks (Table~\ref{tab:param_merge}), it consistently underperforms TIES on the OOD GPTCloneBench across all three seeds for both base models (UX 0.248 $\pm$ 0.050, CB 0.412 $\pm$ 0.059, vs.\ TIES 0.450 and 0.609). PCB falls between the two on OOD as well (UX 0.361, CB 0.567), consistent with its intermediate in-distribution profile. WUDI's per-layer interference-minimization preserves each specialist's decision surface precisely, which helps in-distribution but reduces robustness to AI-generated clones the specialists never saw; TIES's softer sign-election generalizes better OOD. We therefore recommend same-base TIES rather than WUDI as the practical merging recipe (Section~\ref{sec:introduction}, contribution 4) when OOD robustness is a deployment concern.

\noindent Table~\ref{tab:main_results} and Table~\ref{tab:prec_recall} consolidate all results across the three benchmarks. The geometric reasons behind the same-base advantage and the WUDI in-distribution / OOD trade-off are discussed in Section~\ref{sec:discussion}.

\begin{tcolorbox}[before skip=5pt, after skip=5pt, colback=gray!5, colframe=gray!50, 
boxrule=0.5pt, left=4pt, right=4pt, top=3pt, bottom=3pt]
\textbf{Finding 4.} Same-base TIES-merged and architecture-merged models generalize better to unseen GPT-generated clones than any individual encoder specialist, the multi-task baseline, and two zero-shot instruction-tuned code LLMs (up to +35.3\% over the best individual encoder). WUDI, while stronger in-distribution, underperforms on this OOD benchmark, with PCB intermediate, revealing an in-distribution / OOD trade-off within the same-base regime that informs our practical recommendation of TIES.
\end{tcolorbox}

\begin{table*}[!ht]
\centering
\caption{Complete results across all three benchmarks. Three-seed mean $\pm$ std for same-base TIES/WUDI/PCB and cross-base WUDI/PCB.}
\label{tab:main_results}
\begin{tabular}{llcccc}
\toprule
\textbf{Category} & \textbf{Method} & \textbf{BCB F1} & 
\textbf{CLCDSA F1} & \textbf{F1\textsubscript{comb}} & 
\textbf{GPT F1} \\
\midrule
\multirow{4}{*}{Individual} 
& CodeBERT-BCB & 0.894 & 0.464 & 0.679 & 
0.450 \\
& GraphCodeBERT-BCB & 0.891 & 0.495 & 
0.693 & 0.376 \\
& UniXcoder-BCB & 0.940 & 0.269 & 
0.605 & 0.390 \\
& UniXcoder-CLCDSA & 0.259 & 0.924 & 
0.592 & 0.410 \\
\midrule
Multi-task & Combined training (UX) & 0.933 & 
0.922 & \textbf{0.927} & 0.151 \\
\midrule
\multirow{2}{*}{LLM (zero-shot)} & Qwen2.5-Coder-7B-Instruct & --- & --- & --- & 0.454 \\
& DeepSeek-Coder-6.7B-Instruct & --- & --- & --- & 0.120 \\
\midrule
\multirow{9}{*}{Param.\ merge} 
& Cross-base TIES & 0.841 & 0.566 & 0.703 & --- \\
& Cross-base WUDI & .939\tiny{$\pm$.001} & 
.372\tiny{$\pm$.148} & .656\tiny{$\pm$.074} & --- \\
& Cross-base PCB & .935\tiny{$\pm$.004} & 
.542\tiny{$\pm$.122} & .739\tiny{$\pm$.060} & 
.529\tiny{$\pm$.091} \\
& Same-base TIES (UX) & .890\tiny{$\pm$.010} & 
.839\tiny{$\pm$.027} & .865\tiny{$\pm$.012} & 
.450\tiny{$\pm$.042} \\
& Same-base TIES (CB) & .801\tiny{$\pm$.018} & 
.746\tiny{$\pm$.012} & .774\tiny{$\pm$.014} & 
\textbf{.609}\tiny{$\pm$.049} \\
& Same-base WUDI (UX) & \textbf{.935}\tiny{$\pm$.004} & 
\textbf{.863}\tiny{$\pm$.052} & .899\tiny{$\pm$.025} & 
.248\tiny{$\pm$.050} \\
& Same-base WUDI (CB) & .883\tiny{$\pm$.003} & 
.868\tiny{$\pm$.023} & .876\tiny{$\pm$.012} & 
.412\tiny{$\pm$.059} \\
& Same-base PCB (UX) & .910\tiny{$\pm$.030} & 
.854\tiny{$\pm$.040} & .882\tiny{$\pm$.009} & 
.361\tiny{$\pm$.018} \\
& Same-base PCB (CB) & .847\tiny{$\pm$.015} & 
.824\tiny{$\pm$.035} & .836\tiny{$\pm$.014} & 
.567\tiny{$\pm$.003} \\
\midrule
\multirow{2}{*}{Arch.\ merge} 
& CB + GCB & 0.856 & 0.620 & 0.738 & 
0.528 \\
& UX + UX & 0.842 & 0.743 & 0.793 & 0.339 \\
\bottomrule
\end{tabular}
\end{table*}

\begin{table}[!tb]
\centering
\caption{Precision (P) and Recall (R) for key configurations.}
\label{tab:prec_recall}
\resizebox{\columnwidth}{!}{
\begin{tabular}{lcccccc}
\toprule
& \multicolumn{2}{c}{\textbf{BCB}} & 
\multicolumn{2}{c}{\textbf{CLCDSA}} & 
\multicolumn{2}{c}{\textbf{GPT}} \\
\cmidrule(lr){2-3} \cmidrule(lr){4-5} \cmidrule(lr){6-7}
\textbf{Model} & \textbf{P} & \textbf{R} & 
\textbf{P} & \textbf{R} & \textbf{P} & \textbf{R} \\
\midrule
UniX-BCB & .942 & .937 & .563 & .177 & .505 & .318 \\
UniX-CLCDSA & .155 & .788 & .949 & .900 & .483 & .356 \\
Multi-task & .949 & .919 & .947 & .898 & .517 & .089 \\
Same-base TIES & .905 & .896 & .871 & .813 & .516 & .468 \\
Qwen2.5-Coder-7B & --- & --- & --- & --- & .482 & .429 \\
DeepSeek-Coder-6.7B & --- & --- & --- & --- & .647 & .066 \\
Arch.\ (CB+GCB) & .820 & .895 & .509 & .793 & .507 & .551 \\
\bottomrule
\end{tabular}
}
\end{table}

\vspace{-0.20cm}

\section{Discussion}
\label{sec:discussion}

\noindent\textbf{Why Same-Base Merging Works.} \label{sec:why_same_base}

\noindent To explain why same-base merging succeeds while cross-base merging provides only marginal improvement and high variance, we analyze the geometric properties of the corresponding task vectors. We compute the cosine similarity and sign agreement between task vectors from same-base models (UniXcoder-BCB and UniXcoder-CLCDSA) and cross-base models (UniXcoder-BCB and CodeBERT-BCB).

\begin{table}[t]
\centering
\caption{Task vector alignment analysis.}
\label{tab:task_vector}
\begin{tabular}{lrr}
\toprule
\textbf{Metric} & \textbf{Same-Base} & 
\textbf{Cross-Base} \\
\midrule
Cosine similarity & 0.032 & 0.001 \\
Sign agreement & 59.5\% & 50.4\% \\
\bottomrule
\end{tabular}
\end{table}

Table~\ref{tab:task_vector} reveals a stark difference. Same-base task vectors exhibit substantially higher cosine similarity (0.032 vs.\ 0.001), indicating that fine-tuning from the same base produces parameter updates that, while task-specific, remain directionally aligned in parameter space. Cross-base task vectors are nearly orthogonal. Both cosine similarities are small in absolute terms, but the 32$\times$ relative difference is meaningful because TIES merging operates on per-parameter sign agreement, not on global vector alignment. Even modest directional consistency enables productive sign election, as confirmed by the 59.5\% vs.\ 50.4\% sign agreement gap.

The sign agreement is particularly relevant for TIES, which resolves parameter conflicts by electing the majority sign direction. At 59.5\% agreement, same-base task vectors provide a consistent signal for sign election, allowing TIES to identify and retain productive updates from both domains. At 50.4\%, cross-base sign agreement is close to random, leaving TIES with no coherent signal to resolve conflicts. This directly explains the empirical gap observed in Section~\ref{sec:rq2}, where cross-base TIES merging (F1\textsubscript{comb} 0.703) barely improves over individual models, whereas same-base TIES merging achieves 0.865 $\pm$ 0.012 across three seeds.

The geometric argument also explains the behavior of WUDI and PCB under the cross-base setting. WUDI's per-layer objective minimizes the projection of each task vector's residual onto its own subspace (Section~\ref{sec:wudi}). When task vectors reside in different parameter spaces, these subspaces themselves are not aligned, and the per-task projections that WUDI tries to preserve become inconsistent across tasks. Empirically, both cross-base WUDI and cross-base PCB underperform their same-base counterparts on combined F1 and show substantially wider CLCDSA variance across seeds (up to $\pm$0.148, Table~\ref{tab:param_merge}), despite PCB optimizing an unrelated competition-balancing objective. The compatibility requirement therefore does not depend on TIES-specific mechanisms such as sign election, nor on WUDI's interference minimization. It arises from the more fundamental requirement that task vectors reside in a shared parameter space, and it applies to any task-vector merging method that operates under that assumption.

The multi-task model's failure on GPTCloneBench (0.151 F1) provides further insight. Multi-task training optimizes a single set of parameters to jointly minimize loss across both BCB and CLCDSA data, producing a decision boundary tightly fitted to the combined training distribution. When encountering GPT-generated clones, which exhibit substantially different syntactic patterns, this boundary fails entirely. Model merging instead combines two independently optimized specialists through arithmetic combination of their task vectors, preserving the representational diversity of each specialist rather than collapsing it into a single jointly optimized representation. This appears to provide greater robustness to distribution shifts beyond the training domains.

\noindent\textbf{Validation Across Model Families.}
\label{sec:codebert_validation}

\noindent To validate that the same-base principle generalizes beyond UniXcoder, we fine-tune CodeBERT on CLCDSA data and merge with CodeBERT-BCB using TIES under identical settings.

\begin{table}[t]
\centering
\caption{Same-base merging validated on CodeBERT. Merge row is three-seed mean $\pm$ std.}
\label{tab:codebert_validation}
\begin{tabular}{lcccc}
\toprule
\textbf{Model} & \textbf{BCB} & \textbf{CLCDSA} & 
\textbf{F1\textsubscript{comb}} & \textbf{GPT} \\
\midrule
CB-BCB & 0.894 & 0.464 & 0.679 & 0.450 \\
CB-CLCDSA & 0.000 & 0.924 & 0.462 & 0.000 \\
CB TIES merge & .801\tiny{$\pm$.018} & 
.746\tiny{$\pm$.012} & .774\tiny{$\pm$.014} & 
.609\tiny{$\pm$.049} \\
\bottomrule
\end{tabular}
\end{table}

Table~\ref{tab:codebert_validation} confirms that same-base merging works for a second model family. CodeBERT-CLCDSA exhibits complete cross-domain collapse on BCB (0.000 F1), more severe than UniXcoder-CLCDSA (0.259 F1). The merged model nevertheless recovers to 0.801 / 0.746 mean F1 across three seeds, a 14\% combined improvement over the best individual CodeBERT (0.679). The merged CodeBERT also achieves 0.609 $\pm$ 0.049 F1 on GPTCloneBench, the highest zero-shot score across all models in our study.

Prediction distributions reveal why the two CLCDSA specialists differ in cross-domain behavior. CodeBERT-CLCDSA assigns near-zero clone probability to all BCB pairs (mean 0.034), predicting every pair as non-clone with high confidence. UniXcoder-CLCDSA, by contrast, exhibits high uncertainty on BCB pairs (mean clone probability 0.603, std 0.245), over-predicting clones but capturing some true positives in the process. This difference likely reflects UniXcoder's cross-modal AST-based pretraining, which produces more distributed representations that remain partially informative on out-of-domain inputs. Despite these contrasting failure modes, same-base TIES merging recovers cross-domain performance for both model families, confirming that the same-base principle is robust to the underlying model architecture.

\noindent\textbf{Clone-Type Sensitivity Analysis.}
\label{sec:clone_type}

\noindent To verify that merged models detect clones across all difficulty levels, not just easy syntactic clones, we analyze recall by clone similarity range on BigCloneBench. We use token-level Jaccard similarity as a proxy for clone type~\cite{svajlenko2014bigclonebench}, binning positive pairs into four categories that approximate the standard clone taxonomy.

\begin{table}[t]
\centering
\caption{Recall by clone type (approximated via token similarity) on BigCloneBench.}
\label{tab:clone_type}
\resizebox{\columnwidth}{!}{
\begin{tabular}{lccccr}
\toprule
\textbf{Clone Category} & \textbf{UX-BCB} & 
\textbf{UX-CL} & \textbf{TIES (UX)} & 
\textbf{TIES (CB)} & \textbf{Count} \\
\midrule
Type-1/2 ($\geq$0.9) & 1.000 & 0.948 & 1.000 & 
1.000 & 193 \\
Strong T3 (0.7--0.9) & 1.000 & 0.547 & 1.000 & 
1.000 & 53 \\
Moderate T3 (0.5--0.7) & 0.994 & 0.639 & 1.000 & 
1.000 & 155 \\
Weak T3/T4 ($<$0.5) & 0.937 & 0.788 & 0.895 & 
0.900 & 56,419 \\
\midrule
Overall & 0.937 & 0.788 & 0.896 & 0.900 & 
56,820 \\
\bottomrule
\end{tabular}
}
\end{table}

Table~\ref{tab:clone_type} reveals two key findings. First, both merged models achieve perfect recall (1.000) on Type-1/2, Strong Type-3, and Moderate Type-3 clones, matching or exceeding the BCB specialist. Merging does not sacrifice detection of syntactically similar clones. Second, on the dominant and most challenging Weak Type-3/Type-4 semantic clones, merged models achieve 0.895--0.900 recall. While this is slightly below the BCB specialist (0.937), it represents a substantial improvement over the cross-language specialist (0.788), which struggles particularly on higher-similarity clones (0.547 recall on Strong Type-3). The merged model recovers this performance entirely, demonstrating that it integrates complementary detection capabilities across all clone types rather than specializing in only the dominant semantic category.

\noindent\textbf{Complementarity Analysis.}
\label{sec:complementarity}

\begin{figure}[t]
\centering
\includegraphics[scale=0.4]{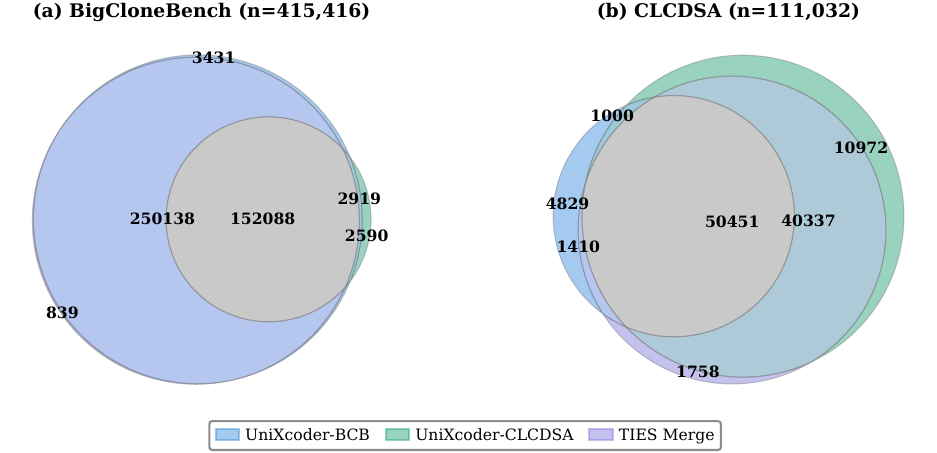}
\caption{Venn diagram of correct predictions per region. Purple regions are samples correctly classified only by the merged model.}
\label{fig:complementarity}
\end{figure}

\noindent To understand whether merging merely interpolates between specialists or captures genuinely complementary knowledge, we analyze per-sample prediction overlap between UniXcoder-BCB, UniXcoder-CLCDSA, and the same-base TIES merge on both test sets (Figure~\ref{fig:complementarity}).

The analysis reveals three key observations. First, the merged model correctly classifies 839 BCB samples and 1,758 CLCDSA samples that \textit{neither} individual specialist could handle. These emergent predictions arise from the combination of complementary task vectors, not from either specialist alone. Second, Cohen's Kappa between the two specialists is near zero ($\kappa = 0.038$ on BCB, $\kappa = 0.045$ on CLCDSA), confirming that they learn fundamentally different decision boundaries. The TIES merge achieves high agreement with each specialist on its home domain ($\kappa = 0.915$ with UniX-BCB on BCB, $\kappa = 0.726$ with UniX-CLCDSA on CLCDSA), demonstrating that merging successfully inherits both boundaries into a single model. Third, all differences between the merged model and individual specialists are statistically significant (McNemar's test, $\chi^2 \geq 2{,}413$, $p < 0.001$ for all comparisons), confirming that the observed improvements are not attributable to chance.

A simple ensemble (averaging softmax outputs) cannot produce these emergent predictions, since if both specialists predict incorrectly their average does as well. The 839 and 1,758 emergent samples therefore demonstrate that merging creates genuinely new decision boundaries rather than interpolating existing ones. Ensembles also incur 2$\times$ inference cost compared to the merged model, which has identical architecture and latency to a single specialist.

\noindent\textbf{Practical Merging Recipe.}
\label{sec:recipe}

\noindent Our findings suggest a concrete recipe for SE teams seeking cross-domain clone detectors. The recipe assumes same-base specialists are already available and joint retraining is impractical; multi-task training remains stronger when all data is jointly accessible.

\noindent\textbf{\textit{(1) Select a base model.}} Choose a single pre-trained code model (e.g., UniXcoder) as the shared base for all domains.

\noindent\textbf{\textit{(2) Fine-tune per domain.}} Create one fine-tuned copy per target domain (e.g., one for same-language, one for cross-language). Standard fine-tuning suffices and no special training procedure is required.

\noindent\textbf{\textit{(3) Merge with TIES.}} Compute task vectors for each specialist and merge using TIES with equal weights ($\lambda_k = 1/K$) and a trim percentile of 20. This step requires no additional training data and completes in under 5 minutes on CPU. We recommend TIES over WUDI for this recipe because, although WUDI achieves higher in-distribution combined F1, TIES generalizes substantially better to out-of-distribution clone types such as AI-generated clones (Section~\ref{sec:rq4}).

\noindent\textbf{\textit{(4) Add new domains incrementally.}} When a new clone type emerges (e.g., AI-generated clones), fine-tune another copy of the same base model on the new domain and re-merge all specialists. No previous training data is needed.

This recipe decouples domain specialization from model deployment, allowing each domain to be handled independently while integration is performed post hoc through merging.

\noindent\textbf{Generalizability Beyond Clone Detection.}
\label{sec:generalizability}

\noindent The same-base merging principle likely extends to other SE tasks where domain specialization creates OOD challenges, such as vulnerability detection across different vulnerability types or code summarization across programming languages. Because our cross-base WUDI and PCB results indicate the compatibility requirement is task-vector geometric rather than method-specific (Section~\ref{sec:why_same_base}), same-base merging should transfer to any task-vector method with compatible specialists. In practice, most organizations standardize on a single base model, making this constraint readily satisfiable.

\section{Related Work}
\label{sec:rltdwork}

\noindent\textbf{Single-Language Clone Detection.} Code clone detection has been studied for over two decades~\cite{roy2007survey, rattan2013software, svajlenko2014bigclonebench, svajlenko2017fast, nakagawa2021nil, jebnoun2022clones, dou2023towards, aversano2007clones,roy2024exploratory}. Early methods used token sequences~\cite{kamiya2002ccfinder}, abstract syntax trees~\cite{baxter1998clone, jiang2007deckard, roy2008nicad}, program dependency graphs~\cite{nguyen2009accurate}, and structural features for Type-1 through Type-3 clones. Deep learning then introduced learned representations~\cite{white2016deep, wei2017supervised, zhang2019novel, wang2020detecting}, and a distinct line targets the harder Type-4 (semantic) clones, where fragments are functionally equivalent but syntactically different, using learned functional similarity~\cite{zhao2018deepsim} and semantic token or graph features~\cite{wu2020scdetector}, evaluated on semantic benchmarks such as SemanticCloneBench~\cite{al2020semanticclonebench}. More recent work advances same-language detection through diverse code representations~\cite{wang2023comparison}, token-based classifiers~\cite{feng2024toma}, and deep subtree interaction~\cite{xu2024dsfm}. Pre-trained code models such as CodeBERT~\cite{feng2020codebert}, GraphCodeBERT~\cite{guo2021graphcodebert}, and UniXcoder~\cite{guo2022unixcoder} achieve state-of-the-art results on BigCloneBench~\cite{svajlenko2014bigclonebench,lu2021codexglue}. These methods improve accuracy within their own benchmarks but remain tied to a single language. We instead combine such strong single-domain detectors into a unified cross-domain model without retraining them.

\noindent\textbf{Cross-Language Clone Detection.} For clones across languages, CLCDSA~\cite{nafi2019clcdsa} uses syntactic features and API documentation, cross-language AST learning~\cite{perez2019cross} operates over tree representations, and C4~\cite{tao2022c4} applies contrastive learning over parallel multilingual data. Li et al.~\cite{li2023zc3} extend this to the zero-shot setting through contrastive learning and cycle consistency. Each of these targets one generalization direction and requires retraining with a modified objective. We instead unify specialist detectors across domains through post-hoc merging, with no retraining and no access to training data at the merging step.

\noindent\textbf{LLM-Based Clone Detection.} A growing body of work studies LLMs for clone detection~\cite{zhang2025exploring, inoue2024improving, li2025nuanced, almatrafi2025code, jiang2026systematic}, generation~\cite{roy2023unveiling}, and analysis~\cite{alam2025classical}, including the construction of LLM-driven benchmarks~\cite{alam2023gptclonebench} and detection via prompting~\cite{dou2023towards}. Concurrent work by Kitsios et al.~\cite{kitsios2025detecting} confirmed that clone detectors suffer significant F1 drops on unseen functionalities and proposed contrastive learning as a mitigation, which reinforces the motivation for a post-hoc combination mechanism. These approaches either fine-tune LLMs for detection or prompt them at inference time, and neither asks whether existing specialist detectors can be unified without retraining. We take the complementary position that compact encoder-based specialists can be merged post-hoc, and Section~\ref{sec:rq4} compares against two zero-shot code LLMs directly. Mixture-of-experts and routing-based methods offer another route to multi-task capability but require architectural modifications and joint training~\cite{fedus2022switch}, placing them outside the post-hoc, training-data-free setting we address. To our knowledge, no prior work has unified cross-domain clone detectors through post-hoc model merging without access to training data, which is the gap our work fills.

\noindent\textbf{Model Merging.} Task arithmetic~\cite{ilharco2022editing} showed that the difference between fine-tuned and pre-trained parameters (the task vector) can be arithmetically combined to edit model behavior. TIES~\cite{yadav2023ties} resolves sign conflicts through trimming and majority-sign election. DARE~\cite{yu2024language} introduces random dropping with rescaling for homologous models. WUDI~\cite{cheng2025wudi} casts task-vector merging as a per-layer optimization that minimizes cross-task interference without training data or rescaling coefficients, and reports state-of-the-art results among task-vector methods on NLP and vision benchmarks. PCB~\cite{du2024parameter} balances parameter competition within and across task vectors to resolve interference without training data. Beyond task-vector methods, Fisher merging~\cite{matena2022merging} uses Fisher information to weight parameter importance, and RegMean~\cite{jin2023regmean} formulates merging as a linear regression problem. Model soups~\cite{wortsman2022model} averages multiple fine-tuned checkpoints to improve robustness in computer vision. HM3~\cite{zhou2024hm3} proposed hierarchical multi-objective merging using reinforcement learning, and merging has been applied to instruction-tuned language models~\cite{yu2024language}, multi-lingual NLP~\cite{muqeeth2024learning}, and domain-specific image classifiers~\cite{ilharco2022editing, wortsman2022model}. These methods were developed and evaluated almost entirely on NLP and vision benchmarks, and none establishes whether the same task-vector compatibility holds for code models that differ in pre-training objectives and tokenizers. We close this gap by evaluating five of these task-vector methods, including the recent WUDI and PCB, directly on cross-domain code clone detection (Section~\ref{sec:rq2}), and by showing that the shared-base condition, rather than the choice of merging algorithm, is what determines success (Section~\ref{sec:why_same_base}).

\noindent\textbf{Multi-Task Learning and Ensembles.} Multi-task learning (MTL) trains a single model on multiple objectives simultaneously, exploiting commonalities between related tasks~\cite{crawshaw2020multi, zhang2023survey}. In software engineering, multi-task approaches have been applied to vulnerability detection with auxiliary objectives~\cite{du2024generalization}. MTL requires joint access to all training data, full retraining when new tasks are added, and careful loss balancing~\cite{crawshaw2020multi}. Ensemble methods combine predictions at inference time~\cite{ganaie2022ensemble} but incur costs scaling linearly with model count. Merging differs from both. Unlike MTL, it operates post-hoc on checkpoints without training data at the merging step, which suits settings where the original data is proprietary, distributed, or expensive to recombine. Unlike ensembles, it produces a single model with no inference overhead. These are not only conceptual differences. We compare merging directly against a multi-task model trained on all data and against inference-time ensembling, and find that merging retains complementary specialist knowledge that multi-task training overwrites, generalizing far better to unseen clone types, while matching a single specialist's inference cost (Section~\ref{sec:rq2}, Section~\ref{sec:complementarity}).

\vspace{-0.3cm}
\section{Threats to Validity}
\label{sec:threats}

\noindent\textbf{Internal validity.} A valid concern is that our greedy layer search may find a suboptimal merge. To address this, we also evaluate global parameter merging, which does not rely on greedy selection, and observe the same same-base advantage. One might also ask whether the results are sensitive to random initialization. We mitigate this by testing every same-base configuration across three random seeds for TIES, WUDI, and PCB. All three methods consistently outperform individual models on in-distribution benchmarks for both UniXcoder and CodeBERT (Table~\ref{tab:main_results}), and although variance on GPTCloneBench is higher, every seed of every same-base TIES configuration exceeds the multi-task baseline. A further concern is that independently trained specialists may differ in learning rates, training epochs, batch sizes, preprocessing strategies, and random data orderings, whereas our specialists use matched configurations. In order to at least partially mitigate this threat, we fix these variables deliberately, so that the merging behavior we observe reflects task-vector compatibility rather than training-quality differences. As partial evidence that the recipe tolerates specialist-level variation, our two model families are trained independently and fail on out-of-domain inputs in contrasting ways, one with high prediction uncertainty and the other with confident misclassification, yet same-base merging recovers both (Section~\ref{sec:codebert_validation}). Of course, further study with more heterogeneous specialists could help establish the limits of the same base recipe under mismatched training conditions.

\noindent\textbf{External validity.} A natural question is whether our findings extend beyond RoBERTa-based encoder models, particularly to decoder-only LLMs. To address this empirically rather than by assertion, we report a direct comparison against two instruction-tuned code LLMs, Qwen2.5-Coder-7B and DeepSeek-Coder-6.7B, in Section~\ref{sec:rq4}. A related risk is over-generalizing from that comparison. We therefore bound it to two models under zero-shot prompting and to the deployment regime we target, namely encoder-based detectors with deterministic predictions at low fixed per-pair cost, and we do not claim that no LLM configuration under few-shot or chain-of-thought prompting could outperform merged detectors. We also note that DeepSeek's low score partly reflects a strong non-clone bias under the one-word prompt rather than an inability to reason about the snippets. It is also possible that our core finding is specific to a single model. We mitigate this by validating the same-base principle on a second model family with contrasting failure modes (Section~\ref{sec:codebert_validation}), since the principle depends on task-vector geometry rather than model architecture or size. Finally, we evaluate only on Java and Java$\leftrightarrow$Python and acknowledge known concerns about BigCloneBench clone type labels~\cite{krinke2022bigclonebench}, while noting that our focus is cross-domain generalization rather than clone-type-specific detection.

\noindent\textbf{Construct validity.} One consideration is that our combined F1 weights both domains equally, which may not match practical priorities. We mitigate this by exposing the merging weights $\lambda_k$, which can be adjusted without retraining to emphasize either domain. Another risk is possible overlap between the training benchmarks (BCB and CLCDSA data), which could introduce data leakage and inflate performance. We address this by reasoning about the direction of the bias. Any overlap would disproportionately benefit the multi-task baseline, which trains jointly on both datasets, whereas our merging approach never accesses the underlying training data, so any leakage would bias results against our method and leave our improvements conservative rather than overstated. One might also question whether the GPTCloneBench evaluation, which relies on 10,000 randomly sampled Java pairs, captures the full diversity of GPT-generated clones. We mitigate this by evaluating across three independent random seeds and observing consistent improvements over all baselines in every run. A final open question is how broadly the in-distribution versus out-of-distribution trade-off across TIES, WUDI, and PCB generalizes.

\vspace{-0.3cm}

\section{Conclusion}
\label{sec:conclusion}
We presented the first systematic empirical study of post-hoc model merging for cross-domain code clone detection. Across five task-vector methods, architecture-level layer stitching, and cross-tokenizer alignment on four code models and three benchmarks, one condition determines whether merging succeeds: specialists must share a pre-trained base. When they do, their task vectors align and combine into a single cross-domain detector that recovers most multi-task performance without training data and generalizes markedly better to unseen AI-generated clones. Different bases leave the vectors near-orthogonal and merging collapses into interference. Because this holds across every method we tested, it reflects task-vector geometry rather than any single algorithm, and yields a simple recipe: fine-tune one shared base per domain and merge with TIES. Several directions follow. The same-base principle should be tested on other SE tasks such as vulnerability detection, defect prediction, and code summarization, where classification-versus-generation architectures may expose method-dependent failure modes. The tokenizer barrier that blocked cross-family alignment invites nonlinear or learned projection approaches. And with sustainability an increasing concern in production deployment, merged detectors offer a low-carbon alternative to per-domain specialist stacks, with inference and storage savings scaling with the number of domains served.
\vspace{-0.3cm}
\section*{Acknowledgements}
This research is supported in part by the Natural Sciences and Engineering Research Council of Canada (NSERC) Discovery Grants program, the Canada Foundation for Innovation's John R. Evans Leaders Fund (CFI-JELF), and by the industry-stream NSERC CREATE in Software Analytics Research (SOAR).

\paragraph{\textbf{Data Availability}}
\label{sec:das}

Our replication package is archived at~\cite{anonymous}. A companion tool paper~\cite{roy2026mergese} describes our web-based interface, available at \url{https://mergese.usask.ca}.

\bibliographystyle{ACM-Reference-Format}
\bibliography{bibliography.bib}

\end{document}